\documentclass[review]{elsarticle}

\usepackage{amsmath}
\usepackage{physics}
\usepackage{verbatim}
\usepackage[utf8]{inputenc}
\usepackage{fancyhdr}

\journal{Optics communications}

\begin{document}

\begin{frontmatter}

\title{Quantum emitter coupled to plasmonic nanotriangle: Spatially dependent emission and thermal mapping}

\author{Adarsh B Vasista}
\author{G V Pavan Kumar\corref{mycorrespondingauthor}}
\address{Photonics and Optical Nanoscopy Laboratory, Division of Physics and Center for Energy Science, Indian Institute of Science Education and Research (IISER), Pune-411008,
India}

\cortext[mycorrespondingauthor]{Corresponding author}
\ead{pavan@iiserpune.ac.in}

\begin{abstract}
Herein we report on our studies of radiative and non-radiative interaction between an individual
quantum emitter and an anisotropic plasmonic nanostructure: a gold nanotriangle. Our theoretical
and three-dimensional electromagnetic simulation studies highlight an interesting connection between : dipole-orientation of the quantum emitter, anisotropy of the plasmonic nanostructure and,
radiative and non-radiative energy transfer processes between the emitter and the plasmonic geometry. For the out of plane orientation of quantum emitter, the total decay rate and non-radiative
decay rate was found to be maximum, showing radiation extraction efficiency of 0.678. Also
the radiative decay rate was greater for the same orientation, and showed a pronounced spatial
dependence with respect to the nanotriangle. Our study has direct implication on two aspects:
designing nanoparticle optical antennas to control emission from individual atoms and molecules
and geometrical control of quenching of emission into plasmonic decay channels.\\
\textcopyright 2016 This manuscript version is made available under the CC-BY-NC-ND 4.0 license http://creativecommons.org/licenses/by-nc-nd/4.0/
\end{abstract}


\end{frontmatter}


\section{Introduction}

Understanding interaction between individual quantum emitters such as atoms and molecules with plasmonic nanostructure has emerged as an important area of research in the context of quantum nanophotonics. By tailoring the geometry of the plasmonic nanostructures, especially by introducing anisotropy in the geometry, one can systematically tune the local density of optical states (LDOS)  \cite{5,23}, which can further effect the radiative and non-radiative processes of a quantum emitter in its vicinity \cite{1,2,4,47}. 
   In the context of quantum emitter interacting with a single anisotropic plasmonic nanostructure, there are two important questions that have emerged in recent times. Firstly, how does the spatial location of the quantum emitter with respect to individual anisotropic plasmonic nanostructure affect its radiative and non-radiative processes? This question is of direct relevance in designing optical antennas  to control atomic and molecular emission\cite{7,27,6,28,24,30,49,51}. With the emergence of nanofabrication methods, accurate placement of single quantum emitters have been achieved\cite{36,37,50}, which may open new avenues in quantum optics and quantum information processing\cite{34,48}. The second question is how the anisotropic plasmonic geometries can be harnessed to effectively control the non-radiative energy transfer between a single quantum emitter and a single anisotropic plasmonic nanostructure. This question has direct implication on the process of quenching of emission by plasmonic channels, and also has connection in designing thermoplasmonic nanoprobes\cite{9,10,25,26,44,46} that are nowadays extensively utilized for applications such as photothermal therapy\cite{38,39,45}, near field sensors \cite{42} etc.

     Motivated by the above-mentioned questions, herein we report on our numerical and theoretical studies of radiative and non-radiative energy transfer processes between an individual quantum emitter and an anisotropy plasmonic geometry - gold nanotriangle. The rationale behind the choice of this geometry was that gold nanotriangle can be nanofabricated by both bottom-up and top-down approaches with excellent control over the geometrical parameters\cite{8,14,15,40,41}. 

In this study, we address the issue of dipole-orientation dependent radiative and non-radiative energy transfer process, and evaluate its spectral and spatial dependence with respect to the plasmonic geometry. We found the out-of-plane dipole (z-polarized) emitter to exhibit greater total decay rates compared to in-plane emitters. Interestingly, the radiative decay rate of  dipole emitters show a strong spatial dependence with respect to the geometry.

\section{Theory}
\subsection{Decay rate of a Quantum emitter}
Under the dipole approximation of particle-field Hamiltonian, one can represent the atom as an oscillating dipole for all practical calculations \cite{5, 52}. The decay rate of a molecular or an atomic emitter can be expressed using Fermi's golden rule, considering weighted sum of all possible decay channels according to which, the total decay rate of an excited emitter is proportional to transition dipole moment \textit{p$_{trans}$}=$\bra{i}\hat{p}\ket{f}$ where $\ket{i}$ and $\ket{f}$ denotes initial and final states of the emitter respectively and $\hat{p}$ represents the dipole operator. When a dipolar emitter is placed in the vicinity of a metallic nanostructure, the emitter will now have two channels to decay. One of those is to directly couple to the far field,$\gamma_{fs}$, and other is to couple to the near field of the structure, $\gamma_{couple}$. Thus coupled power, \textit{P$_{couple}$}, can either get scattered away by the structure to far field or can get absorbed by the structure. Scattered power together with $\gamma_{fs}$ forms the total radiative decay rate. That part of emitted power which ends up in getting absorbed by the structure is a measure of non-radiative decay rate of the emitter. So, in presence of nanostructure we can write the total decay rate of the emitter, $\gamma_{tot}$, as

\begin{equation}
\gamma_{tot}=\gamma_{couple}+\gamma_{fs}
\end{equation}

Where, $\gamma_{couple}, \gamma_{fs}$ are decay rate due to near field coupling with the nanostructure , free space radiative decay rate in the presence of nanostructure respectively. Now, $\gamma_{couple} $ can further be expressed as a sum of absorptive decay rate, $\gamma_{abs}$, (which forms the non-radiative part) and scattered decay rate,$\gamma_{scatt}$, (which forms the radiative part) respectively. The scattered electric field and electric field of the emitter combine to give total radiated power in the far field. Depending upon the phase of the scattered field with respect to to electric field of the dipolar emitter, one can have a destructive interference and hence excite optically dark modes which show complete radiation quench in the far field. 

Now, the total radiation extraction efficiency, (modified quantum efficiency of the emitter in the presence of nanostructure) can be written in terms of $\gamma_{abs}$ and $\gamma_{rad}$  as

\begin{equation}
\eta= \frac{\frac{\gamma_{rad}}{\gamma_{rad}^{0}}}{\frac{\gamma_{abs}}{\gamma_{rad}^{0}}+\frac{\gamma_{rad}}{\gamma_{rad}^{0}}}
\end{equation}

where $\gamma_{rad}^{0}$ is the radiative decay rate of emitter in absence of nanostructure. 

\subsubsection{Radiative and Non-radiative decay rates}
Radiative decay rate for an emitter coupled to nanostructure consists of two parts (i) Direct out-coupling of power by the emitter to the far field (ii) Scattered component of the coupled power by the scatterer. If the dimensions of the scatterer is less than emission wavelength, the scattered power from the structure will be majorly dependent on the induced dipole moment,( $\vec{p}_{induced}$ ),  in the structure.  Using this approximation, normalized radiative decay rate can be represented as \cite{11},

\begin{equation}
\frac{\gamma_{rad}}{\gamma_{rad}^{0}} = \frac{|\vec{p}+\vec{p}_{induced}|^{2}}{|\vec{p}|^{2}} =\frac{P_{rad}}{P_{rad}^{0}}
\end{equation} 

where $\vec{p}=[p_{x},p_{y},p_{z}]$ is dipole moment of the emitter and $P_{rad}^{0}$ is power emitted by dipolar in the absence of nanostructure  .  Exact expression for the radiative decay rate will depend on morphology and polarizability of the structure. 

 Non-radiative decay rate from emitter is quantification of  power absorbed by the nanostructure out of power coupled by emitter to its near field. For an emitter with emission frequency $\omega $ and wavenumber \textit{k}=$\omega$/c , to nanostructure with wavelength dependent dielectric permittivity $\epsilon(\omega)$ it can be expressed as \cite{11},

\begin{equation}
\frac{\gamma_{abs}}{\gamma^{0}_{rad}} = \frac{3}{16} Im\frac{\epsilon(\omega)-1}{\epsilon(\omega)+1} \frac{1}{k^{3}z^{3}} \frac{(p_{x}^{2}+p_{y}^{2}+2p_{z}^{2})}{|\vec{p}|^{2}} =\frac{P_{abs}}{P_{rad}^{0}}
\end{equation}
Neglecting edge and curvature effects equation 4 implies that non-radiant energy transfer is dependent only on the distance, \textit{z}, between emitter and the structure and relative dielectric permittivity $\epsilon(\omega)$ of the structure.

\subsection{Mapping non-radiative transfer of energy}
Energy absorbed by the nanostructure will dissipate mostly as heat. In metallic nanostructures, major source of dissipation is joule heating. So non-radiatively transferred energy can considered, in case of metallic nanostructures, to get dissipated entirely as heat due to joule heating. For far-field illumination, heat power density created will be proportional to absorption cross section of the structure. But, in this case of local excitation, heat power generated will depend on photonic density of states rather than on absorption cross section of the structure. Since the dissipation methodology is the same (joule heating), heat dynamics in the system will be governed by heat diffusion equation with source of heat being electromagnetic power dissipation density in structure due to power emitted by dipole source as,

\begin{equation}
\rho C_{p}(\frac{\partial T}{\partial t}+\textbf{u}_{trans}.\nabla T)+\nabla .(\textbf{q}+\textbf{q}_{r})=-\alpha T:\frac{dS}{dt}+Q
\end{equation}

where $\rho $ is density of material, C$_{p}$ is specific heat capacity at constant pressure, T is absolute temperature, \textbf{u}$_{trans}$ is velocity vector of translational motion, \textbf{q} and \textbf{q}$_{r}$ are heat flux by convection and radiation respectively, $\alpha$ is coefficient of thermal expansion, S is the kirchoff stress tensor and Q is the electromagnetic dissipation power density. For steady state calculations, the time derivatives will vanish from equation 5.

.
With this hindsight, we go on to calculate decay rates of a quantum emitter in the vicinity of an anisotropic nanostructure and quantify the non-radiative energy transfer from the emitter.

\section{Methods}

\subsection{FDTD calculations}
We used Finite Difference Time Domain (FDTD) method to calculate extinction spectrum, near field electric field and normalized decay rates using commercially available solver by Lumerical solutions Inc. The structure under study is a gold nanotriangle of edge length 160nm and thickness of 30nm placed over glass substrate (see figure 1(a)). Triangle was modeled to have rounded corners with corner radius of 20nm to avoid field singularity at corners and also to mimic experimentally realizable object. The area near the triangle was discretized by a non-uniform conformal variant mesh with meshing size of 0.6nm and rest of the simulation area with size of 1nm. Simulation area was terminated by Perfectly Matched Layers (PMLs) to avoid spurious reflections from boundaries. Wavelength dependent dielectric permittivity of gold was taken from experimental details provided by Jhonson and Christy \cite{12} and that of glass (SiO2) from Palik \cite{13}.  For extinction spectrum calculation, a broadband Total Field Scattered Field (TFSF) source\footnote[1]{A TFSF source uses a plane wave illumination and divides simulation area into total and scattering fields} (illumination wavelength, 400nm-1200nm) was used. Absorption and scattering cross sections were calculated using an in-built analysis group in Lumerical FDTD solver and extinction spectrum was calculated in the post processing step as sum of absorption and scattering cross sections. Atomic emitter is modelled as a oscillating classical dipole with current dipole moment 9x10$^{-14}$ Am. It is representative in nature so as to model an excited atom. Decay rates (Radiative, Non-radiative and Total) were calculated using a broadband dipole emitter (emission wavelength, 400nm-1000nm), placed at 5nm above the substrate in two different positions, (i) corner (ii) center of the nanotriangle. Radiative decay rate was calculated by enclosing both emitter and the nanotriangle by a rectangular surface and integrating time averaged Poynting vector over the surface, which gives power outflow from the system, and normalizing this with emission power of the dipole emitter, P$_{0}$.  For the non radiative decay rate calculation, similar procedure was followed with integrating time averaged incoming Poynting vector over a rectangular surface enclosing only nanotriangle, which gives the power inflow to the nanotriangle, followed by subtracting the power scattered off to far field. Intrinsic quantum yield of the emitter is considered to be unity in all calculations. Total decay rate was calculated as sum of its radiative and non-radiative counterparts. 
\subsection{FEM calculations}
We calculated steady state temperature profiles using Finite Element Method (FEM) by solving heat diffusion equation coupled to electromagnetic power dissipation density and surface charge density distribution in commercially available solver by COMSOL Multiphysics 5.2. The structural parameters were maintained same as that of FDTD calculations. Coupled differential equations (Heat diffusion and wave equation) were solved using in-house BiConjugate Gradient Stabilization method (BiCGStab) solver with preconditioning. Simulation area was terminated by scattering boundary conditions and boundaries were kept at a constant temperature of 293.15K. Surface charge density distribution was calculated using Gauss law utilizing local electric field at glass-gold interface. Area around triangle and emitter and emitter was meshed with free tetrahedral mesh of size 0.5nm and relative tolerance of the solver was kept at $10^{-9}$ to ensure accuracy( see section 1, page 1-2 of supplementary information). 

\begin{figure*}[h]
\centering
\includegraphics{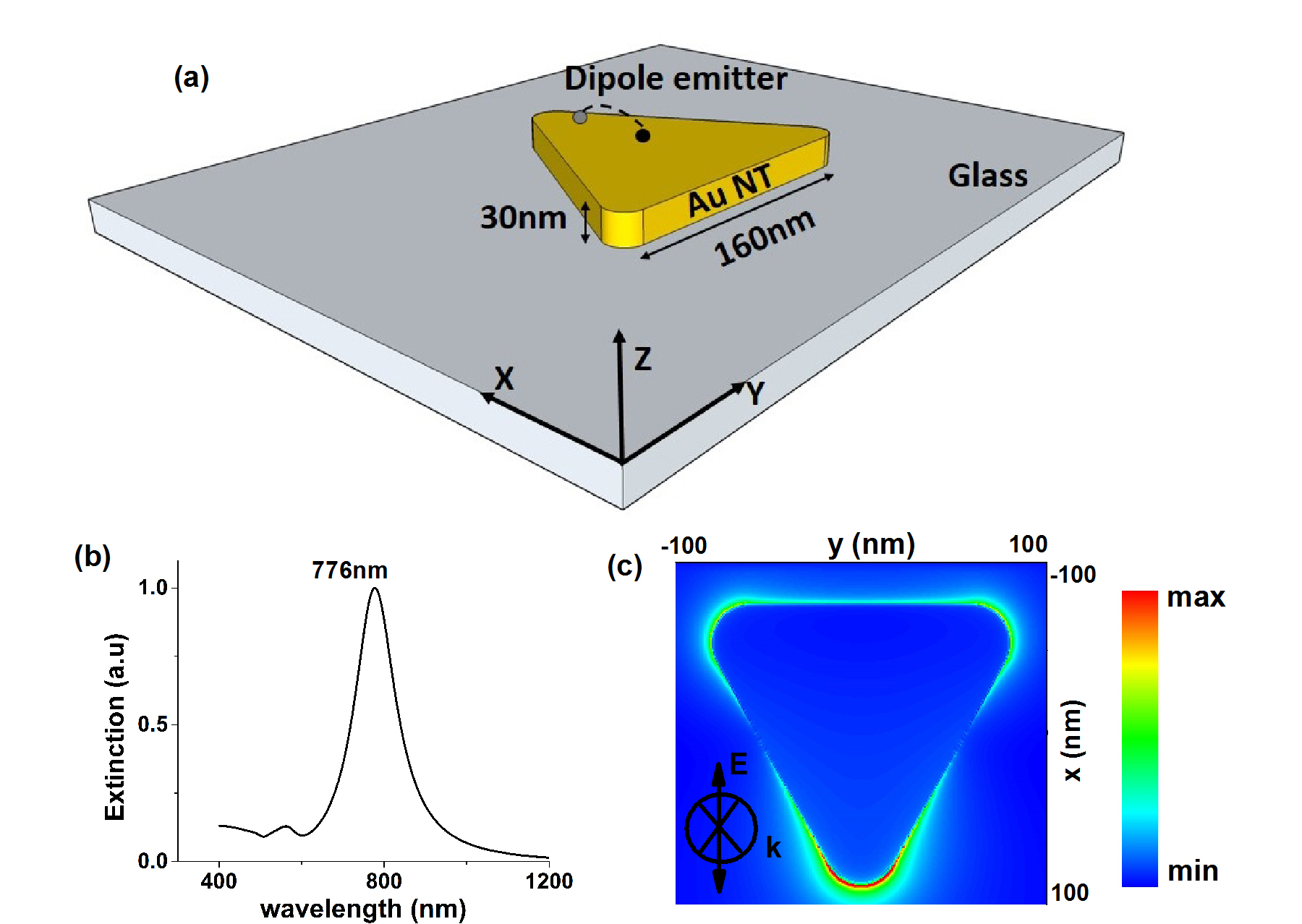}
\caption{(a) Schematic of gold nanotriangle of thickness 30nm and edge length 160nm placed over glass substrate. Calculations were performed by placing a dipole emitter at a distance 5nm above triangle, either at the center or at the corner (represented by grey and black points in the figure). (b) Normalized extinction cross section of the nanotriangle showing a dipolar resonance at 776nm. (c) Near field electric field distribution at the glass-triangle interface for illumination wavelength of 776nm showing the corner mode.The k$-$vector was directed into the plane and the E field polarization is indicated by the double-sided arrow. }
\end{figure*}

\section{Results and Discussion}
\subsection{Plasmon modes of nanotriangle}
 Schematic of the system under study is as shown in figure 1 (a).Typical dimensions of triangle were taken from experimental results of \cite{14}. We have chosen nanotriangles for two major reasons. Firstly, nanotriangles have intrinsic structural anisotropy which shows geometry dependent modes of localization \cite{15}. Depending on the energy of excitation, triangles show various modes of field localization such as corner, edge and center modes. The corner mode, being the lowest order mode, can be excited in visible and near infrared wavelengths. Secondly, electromagnetic resonances of these structures show a critical dependence on edge length and thickness. We have chosen edge length of triangle as 160nm and thickness as 30nm as we limit our discussion to visible wavelength, since triangle with these dimensions is expected show atleast one mode of resonance in Visible and near IR wavelength regime. Calculated extinction cross section of the triangle shown in figure 1 (b) shows a corner mode dipolar peak at 776nm. There is also a weak, broad quadrupolar peak around 540nm. The near field electric field map, calculated at the glass-triangle interface, at 776nm illumination wavelength further clarifies the mode excited in the triangle ( see figure 1(c)) as corner mode. Electric field is localized in the corners of the triangle and field intensity is minimum at the edges and center. Because of strong anisotropy in field localization, emitters placed near such nanostructures experience spatially dependent mode volume, which makes them emit differently depending on the placement with respect to the structure.  

\begin{figure}[h]
\centering
\includegraphics{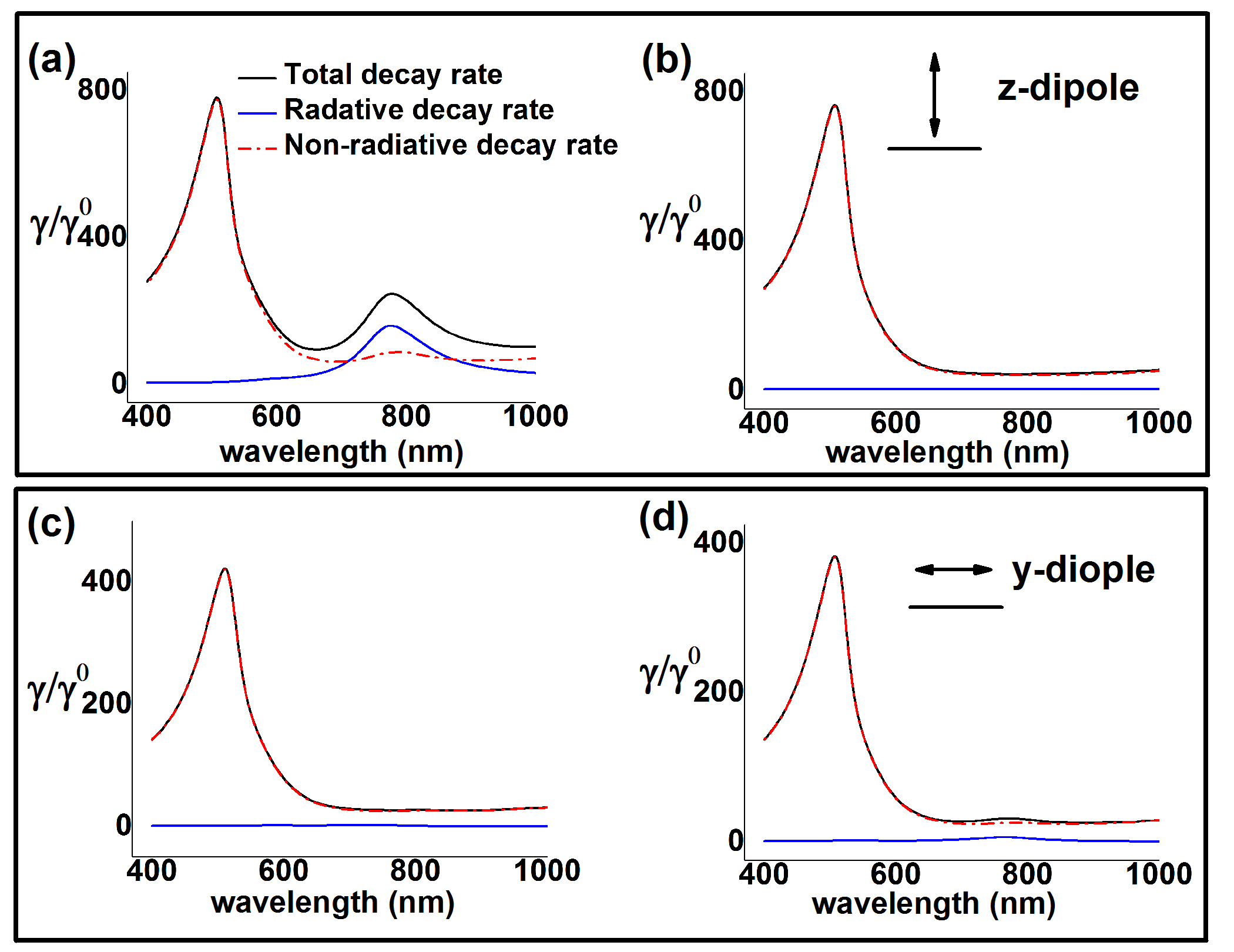}
\caption{Normalized decay rates, $\gamma/\gamma^{0}$ , for z oriented dipole emitter coupled to nanotriangle calculated by placing a dipole at 5nm above (a) corner and (b) center of the triangle. Normalized decay rates, $\gamma/\gamma^{0}$ , for y oriented dipole emitter coupled to nanotriangle calculated by placing it 5nm above (c) corner and (d) center of the triangle. Insets show the orientation of dipoles. }
\end{figure}

\subsection{Calculation of decay rates}
 To  probe effects of strong localized modes, decay rates were calculated by placing a dipole emitter 5nm above triangle in two different positions, above (i) corner and (ii) center of the triangle(see figure 1(a)). Calculated normalized decay rates are as shown in figure 2. The total decay rate curve peaks at two different wavelengths, one can be attributed to radiative decay maximum and another to non-radiative energy transfer.  
\subsection{Radiative decay rate}
For dipole oriented along z-axis, i.e., perpendicular to the surface of the triangle, dipole placed above corner shows a radiative decay peak at 776nm, coinciding with the dipolar corner mode of the structure (see figure 1 (c)), which is absent for the dipole placed over center of the triangle as figure 2 (a) and (b) illustrates. This feature can be explained by the fact that dipole placed above center of the triangle cannot couple to corner mode, hence there will be radiation quenching while the one placed above corner shows radiative enhancement because of coupling with the resonant mode of triangle. Radiation extraction efficiency, $\eta$ ( $\gamma_{rad}$/$\gamma_{tot}$ ), for z-oriented dipole at 776nm is 0.678 for placement of dipole over corner. When dipole orientation is changed to align along y-axis, parallel to the surface of the triangle, there is a total radiation quench in both the placements of emitter as shown in figure 2(c) and (d).  Normalized radiative decay rate for y-polarized dipole placed above center and corner shows considerable quench in emission. This quenching is due to the fact that dipole couples to optically dark quadrupolar mode. (see section 2, page 2-3 of supplementary information)
\subsubsection{Parameters of radiative enhancement}
Radiative enhancement of an emitter coupled system is closely knit  with two important parameters, (i) Spatial modal range of the resonance and (ii) Dipolar moment orientation and positioning of emitter. Because of sharp corners, modal range of triangles are limited and hence will enhance only those emitters which lie in its range and will not affect molecules which are away from the range of the mode. Secondly, dipolar orientation will play a very critical role in exciting certain dark optical modes and hence quenching the emission from emitters even though, emitter is in modal range of the resonance \cite{33}. Structural asymmetry allows to probe both concepts together. Hence, nanotriangles offer an interesting platform to study spatial switching and modulation of fluorescence.      

\subsection{Non-radiative energy transfer}
In all four configurations, there is a strong non-radiative decay peak at 507nm, which can be attributed to the energy transfer by emitter to structure(See figure 2). At this particular emission wavelength, power coupled to the structure, \textit{P$_{couple}$}, is mostly absorbed by the nanostructure and very small amount of power will be radiated to far field resulting in a radiation quench.  As given by equation 4, the non radiative energy transfer is independent of morphology of structure. It depends only on distance between emitter and the nanostructure and relative dielectric permittivity of the structure. For nanostructures made of gold, it occurs around 510nm (Equation 4 qualitatively explains this fact, using wavelength dependent dielectric permittivity). Non-radiative decay rate for dipole orientations parallel to the surface of nanostructure is half as that of orientation perpendicular to structure (see equation 4).  Also, equation 4 neglects curvature and edge effects. This is evident in the case of emitter placed above corner of triangle, where there is a slight change in non-radiative decay rate with respect to the same when placed above center of the triangle in both the dipole orientations. Probing and quantifying non-radiative decay peak is very crucial to understand surface enhanced emissions and also in design of nano sources of heat which act away from their electromagnetic resonances. 

\subsubsection{Quantifying power absorbed by the nanotriangle}
Power which is absorbed by metallic nanostructures will dissipate as heat, mostly due to joule heating \cite{17}. Thus amount of power absorbed by the structure due to near field coupling of an emitter can be quantified using rise in temperature of the structure. For far field illumination, the quantity of heat generated is directly dependent on intensity of impinged light and absorption cross section of the structure. In the case of non-radiative energy transfer, the process of transfer is different i.e., near field coupling and rise in temperature depends on the transition matrix element rather than absorption cross section of the structure. Nevertheless, heat dynamics will remain the same, as it has to do with losses inside the structure such as joule heating. To study the impact of non-radiative transfer on thermal energy of nanotriangle, steady state temperature maps and spatial heat power maps were calculated by placing dipolar source emitting at 507nm wavelength (non-radiative decay peak of the system) above (i) corner and (ii) center and is as shown in figure 3 and 4 respectively. Orientation of dipole is kept along z-axis i.e., perpendicular to surface of triangle. This dipolar orientation has higher non-radiative decay rate than the other two orientations as given by equation 4 and shown in figure 2.
\subsubsection{ Thermal mapping and Heat power calculations}
 Spatial map of temperature is plotted in figure 3 after solving equation 5 in steady state. The plot shows that temperature, at steady state,  is almost uniform inside nanotriangle due to high thermal conductivity of gold \cite{54}. Maximum rise in temperature due to non-radiative near field coupling of the structure was found to be 0.052K for dipole placed over corner and 0.049K for dipole placed over center respectively. Non-radiative decay rate for the dipole placement above corner is slightly more than that for dipole placement over center of the triangle (see figure 2 (a) and (b)). This is the reason behind slight increase of the temperature (see section 3, page 3-4 of supplementary information). 
\begin{figure}[h]
\centering
\includegraphics{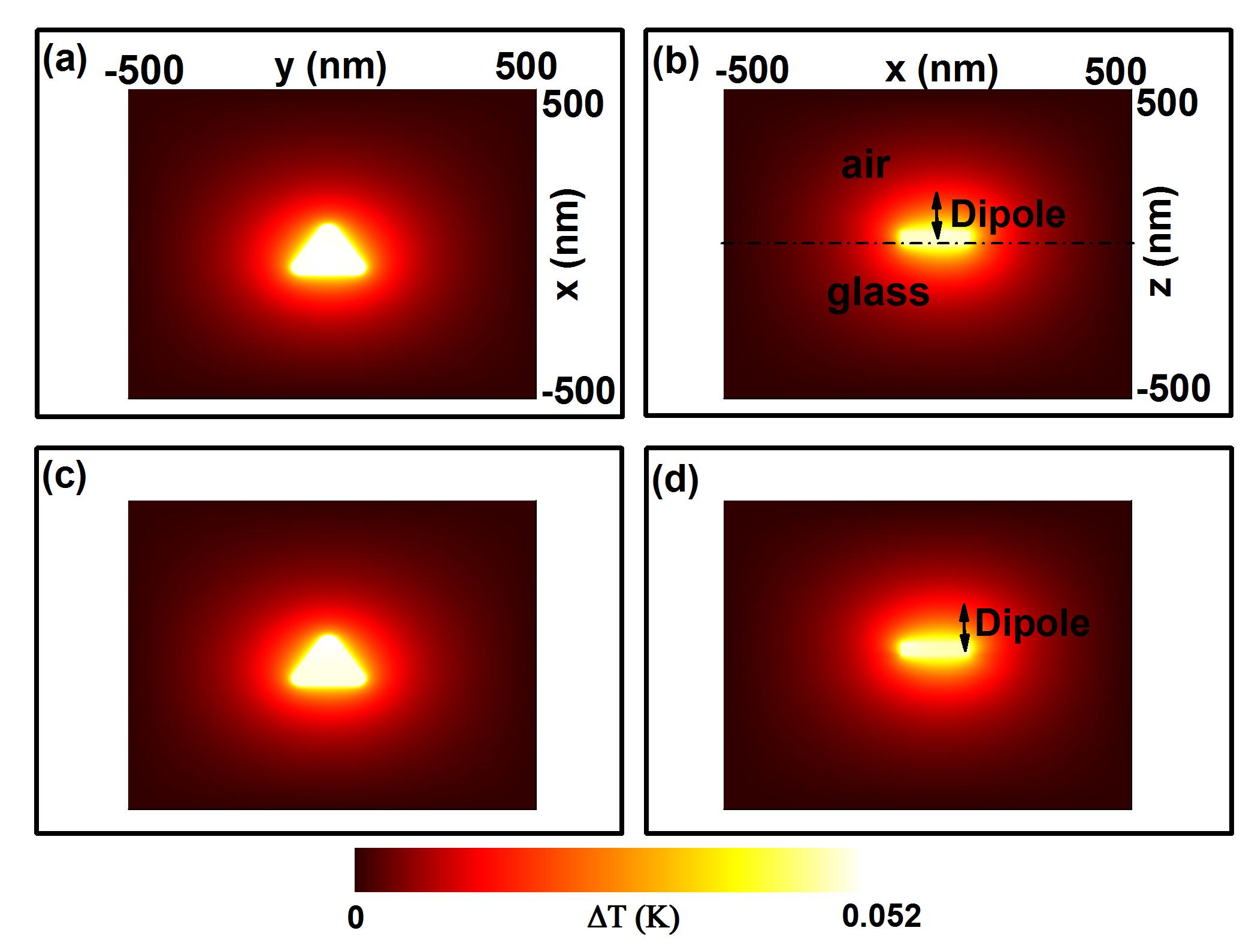}
\caption{ Calculated steady state temperature maps with dipolar emitter, emitting at 507nm, placed, above corner ((a) and (b)) and above center ( (c) and (d) )of the nanotriangle. (a) and (c) are cross sections along x-y plane and (b) and (d) are along y-z plane respectively. Dipole orientation is along z-axis, perpendicular to surface of nanotriangle. Color bar at the bottom represents rise in temperature (in Kelvins).}
\end{figure}

 Heat power density ,$q$, defines the amount of heat absorbed (delivered) by  a system. Basic definition of heat power density remains intact, even though the method of energy transfer is different in this case when compared to far field illumination, as dynamics of heat distribution remain the same. We calculated heat power density for both the configurations, viz., z-oriented dipole above corner and center of the nanotriangle using, \\
\begin{equation}
q(\textbf{r})= \frac{1}{2}Re[\textbf{J}^{*}.\textbf{E}]    
\end{equation}

\begin{figure}[h]
\centering
\includegraphics{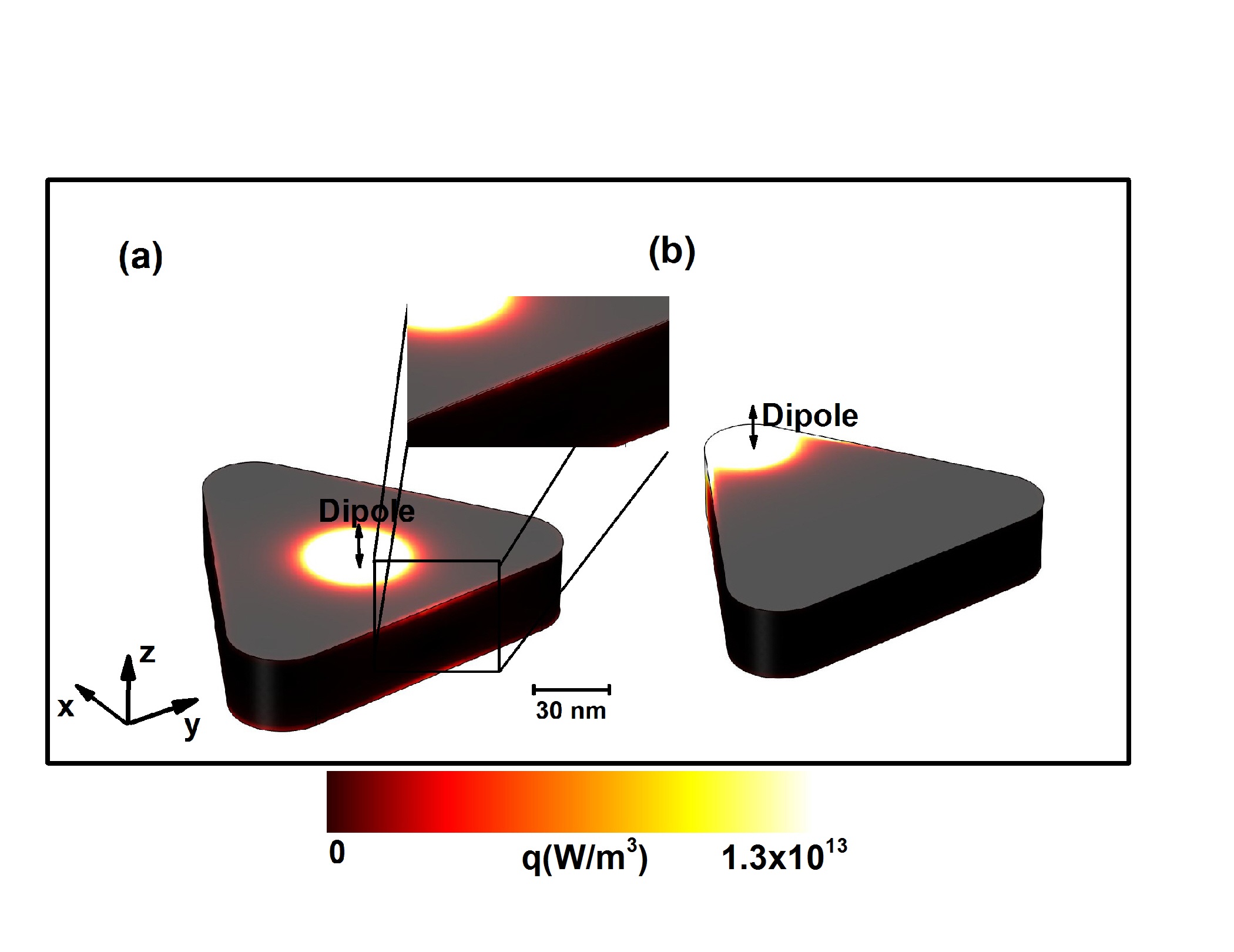}
\caption{ Calculated total heat power density (q) with a z-oriented dipole emitter with emission wavelength 507nm placed 5nm above (a) center and (b) corner of a plasmonic nanotriangle. Single dipolar source is found to dump a maximum of 5.69x$10^{17}$ W/$m^{3}$ when placed over corner and 2.47x$10^{17}$W/$m^{3}$ when placed over center of the nanotriangle respectively . Color bar in the figure represents heat power density in W/$m^{3}$. Any value above 1.3x$10^{13}$ W/$m^{3}$ till 5.69x$10^{17}$ W/$m^{3}$  is represented by white color  }
\end{figure}

  \textbf{J} and \textbf{E} are Current density and Electric field inside the nanotriangle respectively. Integrating equation 6 over the nanotriangle volume gives the total power dumped to the structure (see section 6, page 6 of supplementary information). Even though, steady state temperature is uniform over the structure, heat power generated is not. Figure 4 shows spatial distribution of heat power generated due to the non-radiative transfer of energy from dipolar emitter placed above (a) center and (b) corner, to nanotriangle. Unlike far field focused illumination case \cite{18}, now heat power distribution is dependent on spatial location of emitter with respect to the structure. For dipole placed above corner of triangle (see figure 4 (b) ), the heat power density is localized at corner of the triangle, which is expected because of the high electric field localization at corners and high value of current density at the corner. High value of current density is mostly due to two factors. (i) Spatial location of the emitter with respect to the structure. (ii) Funnel effect of the charges near the corner\cite{53}. Since the emitter is placed right above the corner, electrons near the corner will experience large electric field leading to displacement. Because of the presence of spatial funnel (corner resembles a funnel for flow of electrons) the current density will increase at this location and  hence high heat-power is generated \cite{53}. On the other hand, for dipole placed over center (figure 4(a)) even though heat power is concentrated around the center of the triangle, there is  small portion of power generated at the edges which can be attributed to charge accumulation at the edges and typical distribution of current density due to funnel effect. (see setion 4-5, page 5-6 of supplementary information). Dipole placed over corner of the triangle is found to dump a maximum of heat power density of 5.69x$10^{17}$ W/$m^{3}$  which results in a maximum temperature rise of 0.052K and dipole placed over center dumps a maximum of 2.47x$10^{17}$W/$m^{3}$ which results in a maximum temperature rise of 0.049K.

Quantification of heat power will be very critical to design surface enhanced fluorescence nanoprobes. The macroscopic parameters, heat power together with rise in temperature quantifies non-radiant energy transfer of emitter coupled to a metallic nanostructure. Rise in temperature and hence heat energy in nanotriangle is isolated from electromagnetic resonance of the structure and can be utilized as a potential nano source of heat operating away from resonance of structure.  
\par 
All these calculations are done by considering a dipole in the near field of a nanostructure. But, in experimental situation, there is always a considerable part of heat generated by direct illumination of nanostructure. Heat generated in the nanostructure by direct illumination will depend upon absorption cross section of the system, as previously discussed, and has already been extensively studied \cite{19,20,43}. But in case of nanotriangles, the absorption at 507nm\textendash the non-radiative decay peak, is quite small. Also, in experimental scenario, number of molecules will be large and hence heat power delivered to structure through non-radiative transfer cannot be ignored.  

\section{Conclusion}

Summarizing, we have explored the role of structural anisotropy in radiative and non-radiative emission process of a single quantum emitter coupled to nanotriangle. Full 3D simulations were performed to probe the effects of positioning and orientation of dipolar emitter with respect to nanotriangle on radiative and non-radiative emission process. We have found that an out-of plane dipole placed over corner of the triangle shows maximum radiation extraction efficiency, due to efficient coupling of quantum emitter with dipolar resonance of the structure, showing spatial modulation of emission. We have also quantified the non-radiative energy process using heat power dumped into the nanotriangle. Our results knits effects of geometrical anisotropy with radiative and non-radiative transitions in the quantum emitter. This will have wider relevance not only in understanding light matter interaction at nanoscale, but also in single molecule sensing and spectrally off resonant nanosources of heat.    

\section*{Acknowledgement}
This research was partially funded by DST-Nanomission Grant (SR/NM/NS-1141/2012(G)) and DST Nanoscience Unit Grant (SR/NM/NS-42/2009),India. ABV thanks Mr.Danveer Singh, Mr. Ravi Tripathi and Mr. Arindam Dasgupta for fruitfull discussions. ABV thanks IISER Pune for fellowship.   

\section*{References}

\bibliography{refc}

\end{document}